\documentclass[5p,twocolumn]{elsarticle}

\usepackage{braket}
\usepackage{graphicx}
\usepackage{amsmath}
\usepackage{multirow}

\usepackage{bbm}
\usepackage{textcomp} 
\usepackage{color}

\newcommand{\spinup}{\uparrow}
\newcommand{\spindown}{\downarrow}

\newcommand{\ndag}{{\phantom \dag}}

\begin{document}
\title{A numerical projection technique for large-scale eigenvalue problems}

\author[tugrazitp]{Ralf Gamillscheg}
\ead{ralf.gamillscheg@tugraz.at}

\author[kfmath]{Gundolf Haase}

\author[tugrazitp]{Wolfgang von der Linden}

\address[tugrazitp]{Institute of Theoretical Physics - Computational
  Physics, Graz University of Technology, Graz, Austria}
\address[kfmath]{Institute for Mathematics and Scientific Computing, Karl-Franzens-University, Graz, Austria}

\date{\today}

\begin{abstract}
  We present a new numerical technique to solve large-scale eigenvalue
  problems.  It is based on the projection technique, used in strongly
  correlated quantum many-body systems, where first an effective
  approximate model of smaller complexity is constructed by projecting
  out high energy degrees of freedom and in turn solving the resulting
  model by some standard eigenvalue solver.

  Here we introduce a generalization of this idea, where both steps
  are performed numerically and which in contrast to the standard
  projection technique converges in principle to the exact
  eigenvalues.  This approach is not just applicable to eigenvalue problems encountered in many-body
  systems but also in other areas of research  that result in large scale
  eigenvalue problems for matrices which have, roughly speaking,  mostly a pronounced
  dominant diagonal part.  We will present detailed studies of the
  approach guided by two many-body models.
\end{abstract}

\maketitle

\section{Introduction}

The solution of large, sparse eigenvalue problems is an important task
in engineering, mathematics, and physics, particularly in the field of
strongly correlated quantum many-body systems, such as the high-$T_C$
superconductors, and more recently cold atoms on optical lattices and
coupled light-matter systems.  The treatment of strong correlations
between particles leads to exponentially large eigenvalue problems as
a function of the system size.

To begin with, algebraic eigensolvers like the Lanczos and Arnoldi methods
\cite{Lanczos_EigenvalueProblem,Parlett_Eigenvalue} play an
important role in many body-physics as well as in many other research areas, mainly due to the fact that
they are widely applicable, simple, effective and they yield
numerically exact results. Implementations of the methods can be
obtained from the Internet, e.g. the well-known
ARPACK\footnote{http://www.caam.rice.edu/software/ARPACK/} routine
which includes an implicitly restarting Arnoldi algorithm.  On the downside, only comparably small
systems can be treated by these techniques.   
For most problems in quantum many-body physics, the corresponding geometric  
sizes are much too small.

A couple of sophisticated numerical methods have been developed 
for such systems in
recent years.  The \emph{density matrix renormalization group (DMRG)}
\cite{Schollwoeck_DMRG} is a powerful algorithm for the determination
of ground state properties of large systems, which are not
algebraically feasible. Recently the approach has been embedded into a
wider mathematical framework, the matrix product states (MPS,
\cite{CiracVerstraete_TensorProductStates}). The approach is limited
to 1D or quasi-1D systems. Another method is the \emph{cluster
  perturbation theory} (CPT, \cite{Senechal_CPT}), which constructs
approximations to the Green's function of infinite systems in 1, 2,
and 3 dimensions by exact treatment of small clusters and combining
them by perturbation theory.  An extension of CPT represents the
\emph{variational cluster approach} (VCA, \cite{Potthoff_VCA}) which
improves the results by introducing variational parameters.  An
approach without systematic errors is given by the family of quantum
Monte-Carlo techniques (QMC, \cite{VonDerLinden_QMC}). QMC
simulations, which are based on high dimensional random samples of
some suitable probability-density function have an statistical error which
declines with the sample size. They are applicable to fairly large
systems and to finite temperatures. The drawback of
these methods is the so-called \emph{sign-problem}, that shows up
especially in fermionic models and which makes certain models or
parameter regimes inaccessible \cite{evertz_LOOP}.  
The methods discussed so far are particularly tailored for quantum many-body problems and cannot be applied 
easily, if at all, to matrices of other applications. 

The methods for strongly correlated quantum many-body systems rely on
the assumption that the model has only a few rather  local degrees of
freedom.  Real ab-initio models for strongly correlated quantum
many-body systems are out of reach in any case, and it is inevitable
to reduce the complexity of the  system upon describing the key
physical properties by a few effective degrees of freedom, like in the
multi band Hubbard model \cite{Fulde_ElectronCorrelations}.  In many
cases, these models are still too complicated and cannot be solved
reliably neither analytically nor numerically.  For these cases it has
been proven very useful to construct \emph{effective Hamiltonians}
like the Heisenberg- or $tJ$-model. They are obtained via the
projection technique \cite{Fulde_ElectronCorrelations,Becker} upon
integrating out such basis vectors which correspond to high energy
excitations, or rather which have very large diagonal matrix elements
in a suitable basis.  Of course, it is desirable that the quantitative
results of the effective model are close to those of the original
model. But more than that, it is the qualitative generic physics of
strongly correlated fermions at low energies, which one wants to
understand. Since the original model itself is tailored for that
purpose and is already a crude approximation of the underlying
ab-initio Hamiltonian, it suffices to have an effective model that
still includes the key physical ingredients in order to describe
competing effects of strongly correlated quantum many-body systems.
The standard projection technique defines the space of dynamical
variables (basis vectors) which are of crucial importance for the
low lying eigenvalues, which are marked by small interaction energy
(small diagonal matrix elements). The residual dynamical variables
(basis vectors) are treated in second order perturbation theory. If the
model contains too complicated terms, such as the density assisted
next-nearest neighbor hopping in the $tJ$-model, they are omitted. The
resulting model has a significantly reduced configuration space and is
solved by one of the above mentioned techniques.

Here we present a numerical scheme, that represents a threefold
generalization of the projection technique: a) it combines the two
steps of the construction of the effective model and exact
diagonalization, b) no approximations are made as far as the effective
model is concerned and c) it allows a systemic inclusion of higher
order terms up to the convergence to the exact result.  The projection
step is based on the Schur complement and results in a non-linear
eigenvalue problem, which is solved exactly.

The \emph{numerical projection technique} (NPT), that shall be
discussed in this paper, has several advantages. First of all, it is
not necessary to formulate an explicit effective Hamiltonian, which is
not trivial in more complex models involving several bands and other
degrees of freedom than just charge carriers \cite{Oles}.  And last
but not least, NPT allows to systematically go to higher orders, which
becomes necessary if the coupling strength is merely moderate.

NPT is applicable to other large scale eigenvalue problems as
well. The underlying matrix has to be sparse and the diagonal
elements, after suitable spectral shift, have to fulfill the following
criterion: a few diagonal elements are zero or small and the vast
majority of the diagonal elements is (much) greater than the sum of the 
respective off-diagonal elements.

The method is demonstrated by application to two representative
problems of strongly-correlated quantum many-body physics, the spinless
fermion model with nearest neighbor repulsion and the drosophila of
solid state physicists: the Hubbard model with a local Coulomb
repulsion.  These models are used for benchmark purposes only and it
is not the goal of the present paper to provide a detailed discussion of the fascinating
physics described by these models.

The outline of the paper is as follows. In the second section we
present the two benchmark models. The numerical projection technique
is introduced in detail in the third section and analyzed in the
ensuing sections.
\section{Model systems and basic idea}

The approach that we present below is generally applicable if the
matrix, for which the lowest eigenvalues shall be determined, can be
split into parts of increasing diagonal dominance. What this means in
detail will be clarified using two examples of the realm of many-body
physics, the spinless fermion model with strong nearest neighbor
interaction and the Hubbard model for fermions. Both models are
tailored to study effects of strong correlations of electronic
systems, such as the Mott-insulators \cite{Imada_MetalInsulator}, high
temperature superconductors \cite{Dagotto_CorrelatedElectrons},
manganites \cite{Dagotto_ManganitesReview}, just to name a few of the
very many novel materials with fascinating many-body effects.
The Hamiltonian of the spinless fermion model reads
\begin{equation*}
  \hat H=-t\sum_{\langle i j\rangle}\hat a_{i}^\dag \hat a_{j} +
  V \sum_{\langle i j\rangle}\hat n_i \hat n_j\;,
\end{equation*}
where $\hat a_{i}^\dag$ ($\hat a_{i}^\ndag$) denotes the creation
(annihilation) operator for fermions at site $i$. These operators
have the common fermionic anti-commutator relations (for details see \cite{NegeleOrland}).
The operator $\hat
n_{i}=\hat a_{i}^\dag\hat a_{i}^\ndag$ is the particle number operator
for site $i$. The bracket $\langle i j\rangle$ indicates that the sum
is restricted to nearest-neighbor sites $x_i$ and $x_j$.  There is a
total of $L$ lattice sites $x_i$, which are placed on a simple
cubic lattice in one dimension in the present work.

In the case of the Hubbard model, the spin degree of freedom is also
taken into account and the Hamiltonian reads
\begin{equation*}
  \hat H=-t\sum_{\langle i j\rangle\sigma}\hat a_{i\sigma}^\dag \hat a_{j\sigma} +
  U \sum_i \hat n_{i\spinup} \hat n_{j\spindown}\;.
\end{equation*}
The creation- (annihilation-) operators $\hat a_{i\sigma}^\dag$ ($\hat
a_{i\sigma}^\ndag$) obtain a spin index $\sigma$ with two possible orientations
$\spinup$, $\spindown$. The operator $\hat n_{i\sigma}=\hat
a_{i\sigma}^\dag\hat a_{i\sigma}^\ndag$ is the particle number
operator for spin $\sigma$ at site $i$ and the operator for the
particle density at site $i$ is given by $\hat n_i = \sum_\sigma \hat
n_{i\sigma}$.

The physical behavior of both models is determined by the relative
strength of the hopping parameter $t$ and the two interaction
parameters $U$, which stands for the on-site Coulomb repulsion between
fermions of opposite spin, and $V$, which represents the repulsive
nearest-neighbor interaction. In order to simplify the discussion we
denote both parameters by $V$.  In both cases the most interesting
case is that of (almost) half-filling. That is, the number of
particles in the system is half the maximum capacity, which is $L$ in
the spinless fermion model and $2L$ in the Hubbard model.  Both models
are particularly interesting in the strong coupling case, i.e. for $V
> |t|$.

For both models we use the occupation number basis in real space in
which the interaction part is diagonal and has a value $N_{\kappa} V$,
where $N_\kappa$ is either the number of occupied nearest-neighbor
sites ($N_{n n}$) in case of the spinless fermion model, or the number
of double-occupancies ($N_d$) in case of the Hubbard model. In strong
coupling, states with increasing values of $N_{\kappa}$ are
decreasingly important.  In the projection technique
\cite{Fulde_ElectronCorrelations,Becker} effective strong coupling
models are derived, in which the configuration space is restricted to
the sector $N_{\kappa}=0$, i.e. $N_{n n}=0$ ($N_d=0$) and the
influence of the higher sectors are taken into account up to second
order in $t/V$. A prominent example is the the spin-1/2
antiferromagnetic Heisenberg model which is obtained in the
half-filled case of the Hubbard model or the tJ-model, which is the
generalization away from half-filling. Already in the tJ-model terms
are neglected which belong to the same order in $t/V$ and to the same
$N_d$-sector. In models with more bands and degrees of freedom, the
derivation of an effective strong coupling model can be very
demanding, like e.g. in the spin-orbital model for the manganites
\cite{Oles}.

In this paper, we exploit this idea numerically and generalize it in
such a way, that the influence of high sectors is taken into account
recursively without the need of leaving terms out that in standard
projection technique would complicate the resulting effective
Hamiltonian. For obvious reasons we will refer to this approach as
numerical projection technique (NPT).

\section{Numerical Projection Technique}

\begin{figure}
  \centering
  \includegraphics[width=0.8\linewidth]{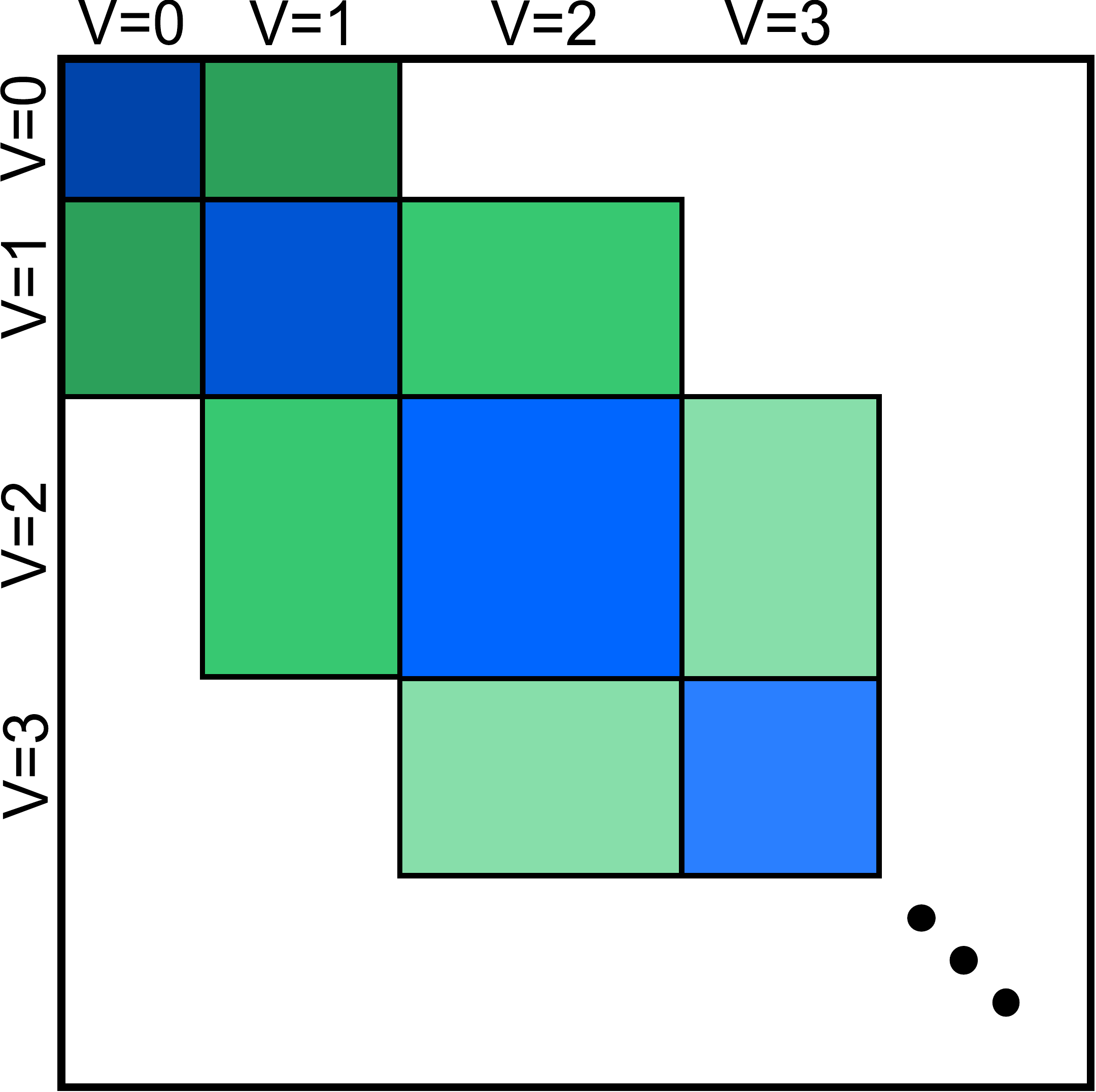}
  \caption{Illustration of the Hamilton matrix of a V-ordered
    occupation number basis. The blue parts contain the hopping terms
    inside a partition and the potential on the diagonal and the green
    parts contain the hopping terms between the partitions.  }
\label{fig:Vordering_matrix}
\end{figure}

In order to exploit the NPT, we reorder the basis vectors according to
$N_{\kappa}$. The corresponding Hamiltonian matrix has a natural block
structure (see Fig. \ref{fig:Vordering_matrix}) corresponding to the
sectors with $N_{\kappa}=0,1,2,\ldots$.

Note, that the hopping of a particle can only change the number
$N_\kappa$ by one. Therefore, only neighboring sectors are coupled and
the matrix has a tridiagonal block-structure.

\subsection{Projection Step}
\label{sec:Vordering_projection}

We start out with a general projection based idea for solving an
eigenvalue problem of a $2\times 2$ block matrix
\begin{equation}
A_0  x =
  \begin{pmatrix}
    A & E\\
    E^\dag & B\\
\end{pmatrix}
\begin{pmatrix}
  x_1\\
  x_2\\
\end{pmatrix}
=\lambda
\begin{pmatrix}
  x_1\\
  x_2\\
\end{pmatrix}\;.
\label{eq:Vordering_matrix_evp}
\end{equation}
The blocks correspond to some suitable partitions of the vector space
under consideration. The sizes of the two partitions shall be denoted
by $p_1$ and $p_2$, respectively. It is not necessary that the two
partitions cover the entire vector space of the original problem.

We assume that $A$ and $E$ are `small' compared to $B$, such that the
lowest eigenvalue of $A_0$ is predominated by the corresponding
eigenvalue of $A$ and the perturbation due to $E$ and $B$ can be
included perturbatively. In order to quantify this idea, we will map
the influence of the second partition into the first partition by a
Schur transformation, similar to the projection technique
\cite{Fulde_ElectronCorrelations,Becker}.  I.e. we multiply
Eq. \ref{eq:Vordering_matrix_evp} from the left with the matrix
\begin{align*}
    \begin{pmatrix}
    I & -EB^{-1}\\
    0 & I\\
\end{pmatrix}
\end{align*}
resulting  in
\begin{equation*}
  \begin{pmatrix}
    S & 0\\
    E^\dag & B\\
\end{pmatrix}
\begin{pmatrix}
  x_1\\
  x_2\\
\end{pmatrix}
=\lambda
  \begin{pmatrix}
    I & -EB^{-1}\\
    0 & I\\
\end{pmatrix}
\begin{pmatrix}
  x_1\\
  x_2\\
\end{pmatrix}\;,
\end{equation*}
with $S=A-EB^{-1}E^\dag$ being the Schur-complement. The second line
of the eigenvalue equation yields
\begin{equation}
\label{eq:Vordering_proj_x2}
  x_2=-(B-\lambda)^{-1}E^\dag x_1\;.
\end{equation}

Inserting $x_2$ into the first line leads to
\begin{equation}
\label{eq:Vordering_proj_x1}
  S_\lambda x_1:=(A-E (B-\lambda)^{-1} E^\dag)x_1=\lambda x_1\;.
\end{equation}

Note that this equation is in principle exact, no approximations have
been made so far. The original problem
(Eq. \ref{eq:Vordering_matrix_evp}) has thus been projected into the
subspace of the first partition which is of smaller size, but at the
prize of a non-linear eigenvalue problem.  By the above procedure we
obtain only those eigenvectors $x=(x_1,x_2)$ of the $2\times 2$ block
matrix with non-vanishing $x_1$, i.e. those vectors which evolve
perturbatively from those of $A$. There are additional eigenvectors of
the form $x=(0,x_2)$, which can, however, be omitted, since their
eigenvalues are of order $O(V)$ instead of $O(t)$.

\subsection{Solution of the non-linear eigenvalue problem}
\label{sec:Vordering_proj_series}

In the numerical projection technique, to be outlined below, the
second partition will consist of a single sector. Hence $B$ will be
the sub-matrix of the Hamiltonian corresponding to basis vectors of that particular sector
and it will consist of a kinetic term and an interaction term, $B =
\tilde B + B_V$, with $B_V = N_\kappa V \bf{I}$, where a $\bf{I}$ is a
unit matrix.  By virtue of this structure and the fact that $|V|\gg
|t|$, the numerical solution of the non-linear problem can be
simplified significantly.  We can expand the inverse in powers of
$B_V^{-1}$ or rather $(B_V-\lambda)^{-1}$. The latter step is
justified, since the lowest eigenvalues are of order $t$ plus
corrections of order $V^{-1}$, hence $B_V-\lambda = O(V)$. The
expansion yields
\begin{eqnarray*}
  \frac{1}{B-\lambda} &=&  \frac{1}{B_V-\lambda + \tilde B}\\
 &=& \frac{1}{B_V-\lambda} - \frac{1}{(B_V-\lambda)^2}\tilde B +\dots \\
&=&\sum_{\nu=0}^\infty \frac{1}{(B_V-\lambda)^{\nu+1}} (-\tilde B)^\nu \;,
\end{eqnarray*}
and the calculation of $S_\lambda$ becomes
\begin{equation}
  \label{eq:Vordering_proj_Sexpansion}
  S_\lambda=A+\sum_{\nu=0}^\infty (\lambda-B_V)^{-(\nu+1)} \;E\;{\tilde B}^\nu \;E^\dag \;.
\end{equation}

Note that the expressions $(\lambda-B_V)^{-(\nu+1)}$ are only
numbers. For an iterative solution of the non-linear eigenvalue
problem, $S_\lambda$ has to be computed repeatedly for different
values of $\lambda$, but only the pre-factors
$(\lambda-B_V)^{-(\nu+1)}$ are modified.  The matrices $E \tilde B^\nu
E^\dag$, however, are independent of $\lambda$. Therefore, they can be
calculated once and stored, since they are in general of moderate
size, as will be discussed later on.

This expansion can also be used in Eq. \ref{eq:Vordering_proj_x2} in
order to evaluate the second part of the eigenvector, i.e.
\begin{equation*}
 x_2 = -(B-\lambda)^{-1}E^\dag=-\sum_{\nu=0}^\infty (\lambda-B_V)^{-(\nu+1)} \tilde B^\nu E^\dag\;.
\end{equation*}

The non-linear eigenvalue problem can now be solved iteratively, by
using an initial approximation for $\lambda$ to build $S_\lambda$,
then solve the eigenvalue problem by suitable means and use the
resulting eigenvalue as a new approximation for $\lambda$. Let
$x_\lambda$ be the normalized eigenvector of $S_\lambda$ to the lowest
eigenvalue $\mathcal E(\lambda)$. In the next recursion
$\lambda^{\text{(new)}}= \mathcal E(\lambda)$ is used as new value for
the parameter $\lambda$.

A more sophisticated approach to obtain an improved value for
$\lambda$ is provided by the Newton-Raphson method.  We are actually
seeking the roots of $\Phi(\lambda)=\mathcal E(\lambda)-\lambda$. The
Newton-Raphson approach yields
\begin{equation*}
  \lambda^{\text{(new)}}=\lambda-\frac{\Phi(\lambda)}{\Phi'(\lambda)}
=-\frac{1}{\mathcal E'(\lambda)-1} \mathcal E(\lambda) + \frac{\mathcal E'(\lambda)}{\mathcal E'(\lambda)-1}  \lambda\;.
\end{equation*}
The expression shows that this can be transformed to an equation of
form $\lambda^{\text{(new)}}=\alpha \mathcal E(\lambda)+(1-\alpha)\lambda$ with
\begin{equation*}
  \alpha=-\frac{1}{\mathcal E'(\lambda)-1}\;.
\end{equation*}
Note that in the simple iteration scheme $\alpha=1$.
In order to calculate $\mathcal E'(\lambda)$ we can use the Hellmann-Feynman theorem resulting in
\begin{align*}
    \mathcal E'(\lambda) &= x_\lambda^\dag S_\lambda' x_\lambda\,.
\end{align*}
The derivative of $S_\lambda$
(Eq. \ref{eq:Vordering_proj_x1}) with respect to $\lambda$ therefore reads
\begin{equation}
  \label{eq:Vordering_proj_Sprime}
  S_\lambda'=- E (B-\lambda)^{-2}  E^\dag\;,
\end{equation}
which, like Eq. \ref{eq:Vordering_proj_Sexpansion}, can be expressed
in a Taylor expansion
\begin{equation}
  \label{eq:Vordering_proj_Sprime_expansion}
 \mathcal E'(\lambda)= -\sum_{\nu=0}^\infty (\nu+1)(\lambda-B_V)^{-(\nu+2)}  x_\lambda^\dag E  \tilde B^\nu E^\dag x_\lambda\;.
\end{equation}

As before, the stored matrices $E \tilde B^\nu E^\dag$ can be reused.

\subsection{Iterative inclusion of higher sectors}

Now we are in the position to describe the numerical projection
technique.  To begin with, we consider the following two partitions: the first partition may contain the
lowest sectors $N_\kappa = 0,1,\ldots,N'$. The second partition consists of merely one sector, namely 
$N_\kappa = N'+1$. The sizes of
the partitions are $p_1$ and $p_2$, respectively.  The projection
step, discussed before, is used to combine the two partitions. This step can now be invoked repeatedly 
in order to include increasingly higher sectors.

\begin{enumerate}
\item[a)] We begin with the first partition, solve the eigenvalue problem for $A_1$  and determine the corresponding
  unitary matrix $V_1$ containing the corresponding $p_1$ eigenvectors.

\item[b)] The matrix for the first two partitions has the following block structure
  \begin{equation*}
  A_2=
  \begin{pmatrix}
    A_1 & E_1\\
    E_1^\dag & B_1\\
  \end{pmatrix}\;.
\end{equation*}

\item[c)] Next we perform a unitary transformation on $A_{2}$ provided by the unitary matrix   $\text{diag}(V_1,I)$ resulting in 
\begin{equation}
\tilde A_2:=
    \begin{pmatrix}
    D_1 & V_1^\dag E_1\\
    E_1^\dag V_1 & B_1\\
  \end{pmatrix}
\label{eq:Vordering_iteration_concept}
\end{equation}
with $D_1:=V_1^\dag A_1 V_1$ being a $p_1$-dimensional diagonal matrix.

\item[d)] Now we solve the eigenvalue problem for \hbox{$\tilde A_2
    \tilde V_2= \tilde V_2 D_2$} as outlined in the previous section
  and construct the eigenvector matrix for the entire matrix $A_2$:
\begin{equation*}
  V_2=
  \begin{pmatrix}
    V_1 & \\
    & I\\
  \end{pmatrix}
  \tilde V_2\;,
\end{equation*}
with $I$ being a identity matrix of appropriate size. As discussed before
$\tilde V_2$ only contains the lowest $p_1$
eigenvectors. Consequently, $D_2$ is again a $p_1$-dimensional
diagonal matrix.

\item[e)] Finally we include the blocks $E_2$ and $B_2$ of the next
  sector and perform again a unitary transformation with
  $\text{diag}(V_2,I)$ as in
  Eq. \ref{eq:Vordering_iteration_concept}\, leading to
  \begin{equation*}
  \tilde A_3 =
    \begin{pmatrix}
      D_2 & V_2^\dag E_2^\ndag\\
      E_2^\dag V_2^\ndag & B_2\\
    \end{pmatrix}\;.
  \end{equation*}
\end{enumerate}
The recursion steps (d) and (e) are continued until the desired
accuracy is reached.

\subsection{Truncation of the first partition}

The advantage of the numerical projection technique so far is the fact
that (non-linear) eigenvalue problems have to be solved for matrices
of a size given by the first partition $p_1$, which is obviously
smaller then the original size.

\begin{figure}
\centering
\begin{minipage}[c]{0.28\linewidth}
  \centering
  \includegraphics[width=\linewidth]{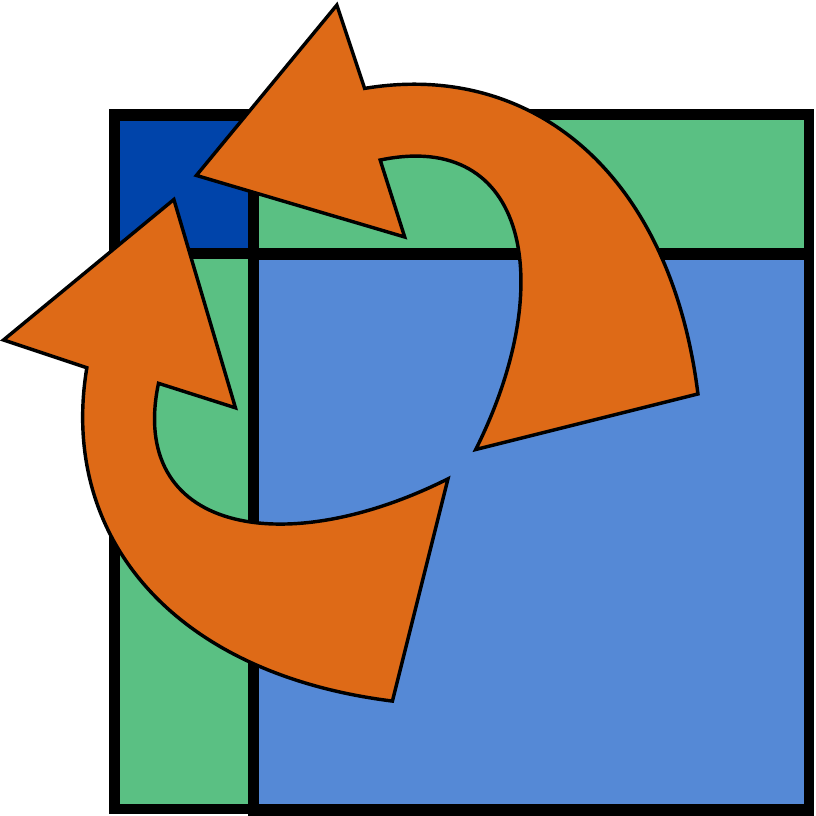}
\end{minipage}
\begin{minipage}[c]{0.03\linewidth}
  \centering \Huge \textrightarrow
\end{minipage}
\begin{minipage}[c]{0.28\linewidth}
  \centering
  \includegraphics[width=\linewidth]{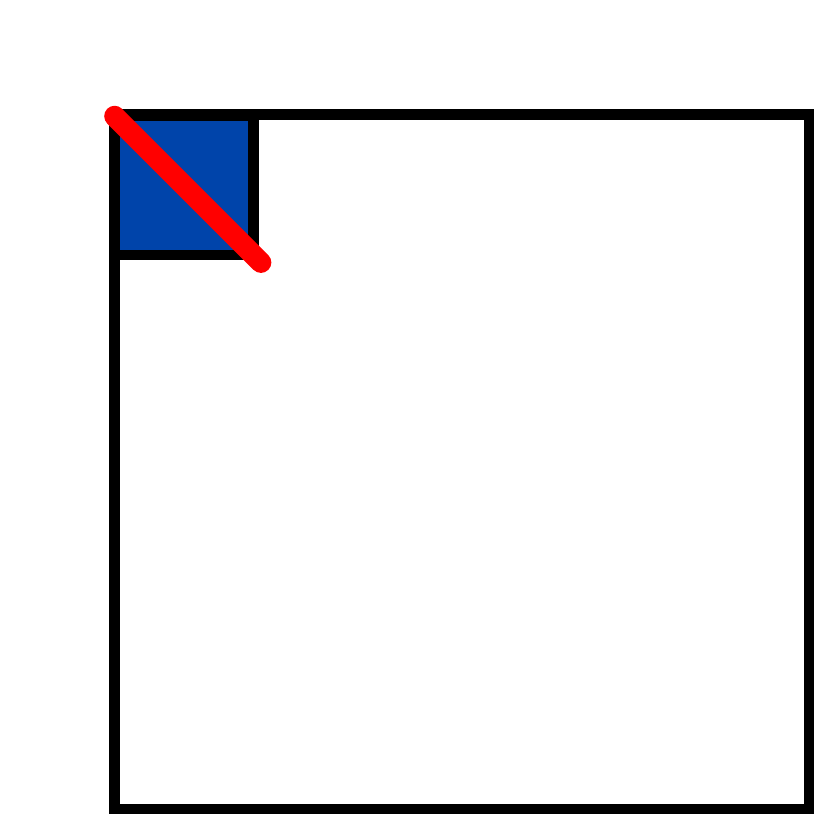}
\end{minipage}
\begin{minipage}[c]{0.03\linewidth}
  \centering \Huge \textrightarrow
\end{minipage}
\begin{minipage}[c]{0.28\linewidth}
  \centering
  \includegraphics[width=\linewidth]{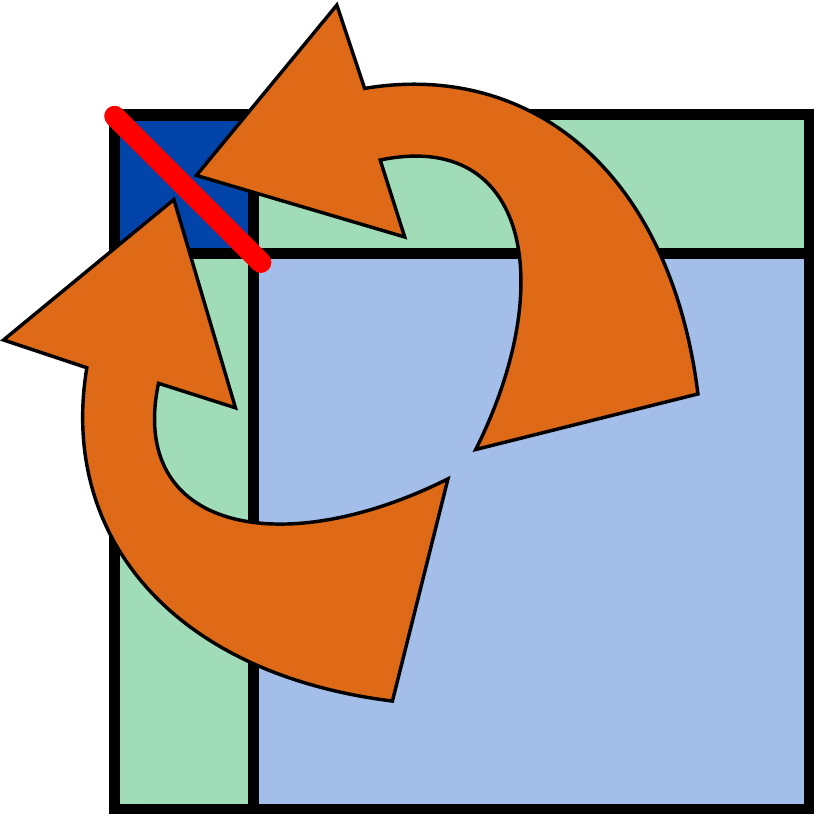}
\end{minipage}
\begin{minipage}[c]{0.03\linewidth}
  \centering \Huge ...
\end{minipage}
\caption{Truncation procedure for the scheme presented in the
  text. Each partition is consecutively projected onto the original
  effective matrix and diagonalized.}
\label{fig:Vordering_flapping}
\end{figure}

It may happen that the first partition, even if it covers merely the
first sector, is still very large. This is the case in the Hubbard
model, where already the lowest sector for $N_\kappa =0$ yields 
\hbox{$p_{1}={L\choose N_\uparrow}{L-N_\uparrow \choose N_\downarrow}$}, which at
half-filling becomes $p_{1} ={L \choose L/2}$. Next we will try and reduce the
size of the first partition.  Firstly, we can exploit translational
invariance and solve the eigenvalue problem in the individual $k$-space
sectors.  We can furthermore restrict the number of vectors, which are
retained in the first partition, based on some suitable criterion as to
there importance for the lowest eigenvalue.  We will use only the most
obvious and simple criterion, namely to keep those vectors, which belong to
the lowest eigenvalues $D_2$ of $A_1$.  More sophisticated criteria as
in DMRG are also conceivable.

We will use a predefined value $\tilde p_1\le p_1$ for the truncation
size of the first partition. That means, we solve the eigenvalue
problem of $A_1$ of size $p_1\times p_1$ and retain only those $\tilde
p_1$ eigenvectors in $V_1$ which correspond to the lowest
eigenvalues.  According to the recursion procedure, in the following
recursion steps the diagonal matrices entering the left-upper block
$D_n$ will all have the size $\tilde p_1$ and henceforth only
(non-linear) eigenvalue problems of this size occur.

\section{Numerical experiments}

We begin with the spinless fermion model with periodic boundary
condition (pbc) at half-filling.  Due to the conservation of the total
momentum $k$ only those basis vectors with the same $k$ need to be
taken into account.  For the spinless fermion model at half-filling
the ground state is degenerate and has $k=\pm \frac{\pi}{2}$. We will
therefore only take basis vectors into account, that have a total
momentum $k=\pm \frac{\pi}{2}$.\footnote{We keep both, to allow real
  valued matrices and eigenvectors.} The selection of the total
momentum leads to a significant reduction of the computational
effort. On the one hand, the size $p_1$ of the initial partition is
reduced roughly by $L$ and on the other hand it can be avoided that in
the truncation step vectors are retained, that are irrelevant for the
ground state of the original system.

To begin with,
we choose as first partition the first sector with $N_\kappa=0$,
which consists of merely 2 states with particles on every other site
(charge density wave), i.e. $\tilde p_1 =p_1 = 2$. These two vectors are the real valued linear combinations
of the $\pm \frac{\pi}{2}$ basis vectors.
\begin{table}[h]
  \centering 
  \begin{tabular}{lccrcc}
    \hline
    \hline
       $L$ & M &$t_\text{NPT}$ [s]& $t_\text{AR}$ [s]& $E_\text{NPT}$ & $\left|\frac{E_\text{NPT}-E_{0}}{E_0}\right|$\\
    \hline
    16&12870    &  0.0   &  0.2& -1.57    & 0.007 \\
    18&48620    &  0.0   &  0.8& -1.76    & 0.010 \\
    20&184756   &  0.1   &  4.1& -1.95    & 0.012 \\
    22&705432   &  0.3   & 10.4& -2.14    & 0.014 \\
    24&2704156  &  0.6   & 87.1& -2.33    & 0.014 \\
    \hline
    \hline
  \end{tabular}
  \caption{Run-time and accuracy comparison between NPT ($t_\text{NPT}$, $E_\text{NPT}$) and traditional Arnoldi eigensolver (ARPACK, $t_\text{AR}$, $E_0$ ) for
    the spinless fermion model ($|V/t|=10$) at half-filling for different system sizes and pbc. The recursion was stopped at an absolute accuracy of $10^{-4}$. The matrices have size $M\times M$. All computations in this paper have been performed on a computer with Intel Core 2 Duo 3GHz,  2GB RAM.}
  \label{tab:Vordering_1}
\end{table}

Tab. \ref{tab:Vordering_1} exhibits a comparison of the run time for
various system sizes. It can be seen that NPT is significantly faster
than the highly optimized ARPACK routine (Arnoldi implementation) for
increasing system size. We observe that even for $10^{6}\times 10^{6}$-matrices
the CPU time is less than 1 sec. and more than a hundred times faster than the highly optimized
ARPACK code.

\subsection{Dependence on the size of the initial partition}

\begin{figure}[h]
  \centering
   \includegraphics[width=0.8\linewidth]{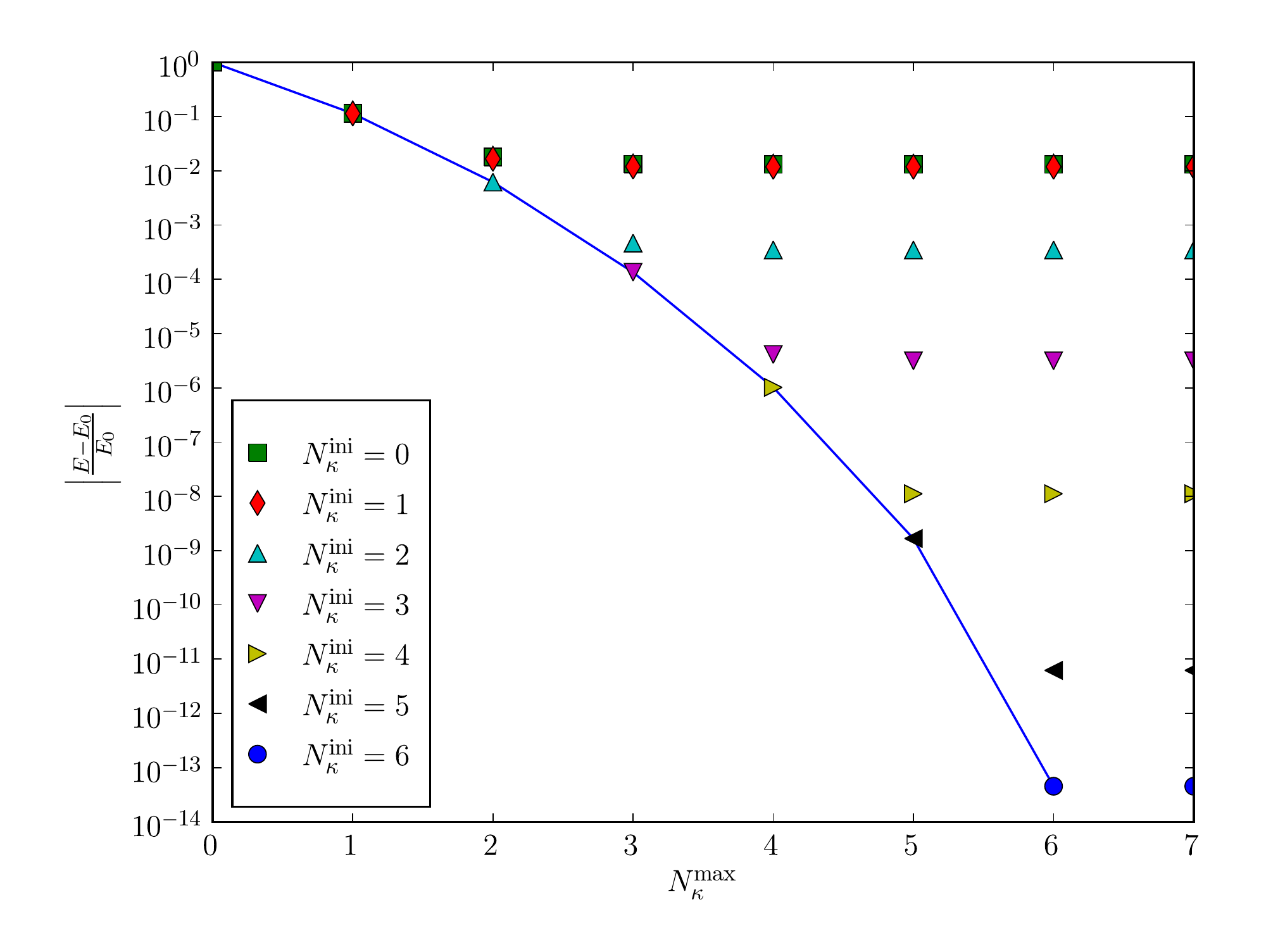}
   \caption{Convergence of the lowest eigenvalue with the largest sector ($N_\kappa^\text{max}$)
   included in the recursion,
    depending on the initial partition. In all cases, $\tilde p_1 =2$
     was used, i.e. the 2 lowest
     eigenvectors of the initial partition where kept. The blue solid line
     indicates the exact values corresponding the various subspaces (see text). The results are compared with the exact
     eigenvalue $E_0$ of the original matrix on a logarithmic scale. System parameters are $L=20,N=10,|V/t|=10$.}
  \label{fig:Vordering_starting_part}
\end{figure}

In Fig. \ref{fig:Vordering_starting_part} the upper curve (green
squares) corresponds to the case that the initial partition consists
of the first sector ($N_\kappa=0$) only, with $\tilde p_1=p_1 =2$ and
shows the dependence of the lowest eigenvalue on the maximum number of
sectors, specified by $N_\kappa^\text{max}$, included in the
recursion. We observe that the recursion rapidly levels off at a
moderate relative accuracy of 1 percent.  The reason is due to the
fact, that the vectors that are omitted during the recursion
procedure, are relevant for a higher accuracy.

One way to increase the accuracy is to choose a greater initial
partition upon including all sectors up to $N_\kappa^\text{ini}$, but
still using a truncation size of the first partition $\tilde p_1 = 2$.
The results are also depicted in
Fig. \ref{fig:Vordering_starting_part}. The turquoise upward
triangles, e.g., depict the results for $N_\kappa^\text{ini}=2$,
i.e. the initial partition contains sectors up to $N_\kappa = 2$ and
the subsequent recursions mix in $N_\kappa
=3,\ldots,N_\kappa^\text{max}$. The relative accuracy increases by one
order of magnitude but is still restricted. In all curves we observe
that the result is converged when two more sectors are included beyond
the ones present in the initial partition, which is  obviously a
consequence of the small truncation size $\tilde p_1 =2$.  The blue
solid line indicates the result of the lowest eigenvalue obtained in
the subspace spanned by the vectors of the sectors
$N_\kappa=0,\ldots,N_\kappa^\text{max}$.

\subsection{Dependence on the truncation size}

So far we have seen that the accuracy is limited when the truncation
size is restricted to 2, even if the size of the initial partition is
increased and all sectors are iteratively included. One is therefore
driven to study the dependence on the truncation size. To this end we
consider one particular system ($L=20,N=10, |V/t|=10$) and choose as
initial partition the lowest two sectors with $N_\kappa=0$ and
$N_\kappa=1$. The size of the initial partition is $p_1=182$.  In
Tab. \ref{tab:Vordering_number_init} the energies are listed for
different truncation sizes $\tilde p_{1}$. In all cases NPT is iterated until an
absolute accuracy of $\epsilon_\lambda=10^{-4}$ is reached. We observe
a clear improvement as compared to the case $\tilde p_1=2$, although the
accuracy is certainly limited by the fact, that only the first two
sectors are included in the initial partition. We find that the
accuracy achieved by using three sectors can never be reached.
\begin{table}[h]
  \centering
  \begin{tabular}{rccr}
    \hline
    \hline
      $\tilde p_1$ &  $E_\text{NPT}$ & $\left |\frac{E_\text{NPT}-E_0}{E_0}\right|$ & $t_\text{iters.} [s]$\\
    \hline
       2 & -1.9565 & 0.012 &   0.66 \\
       4 & -1.9573 & 0.011 &   1.34 \\
       6 & -1.9594 & 0.010 &   2.25 \\
      10 & -1.9629 & 0.008 &   3.76 \\
      14 & -1.9713 & 0.004 &   6.79 \\
      16 & -1.9749 & 0.003 &   9.21 \\
      \hline
      3 sectors & -1.9793 &0.001\\
      exact& -1.9800 &  \\
    \hline
    \hline
  \end{tabular}
  \caption{Dependence of the lowest eigenvalue on the truncation size $\tilde p_1$ for a system with $L=20$, $N=10$, $|V/t|=10$, $\epsilon_\lambda=10^{-4}$. Here, the initial partition consists of the first two sectors.
    In addition the exact lowest eigenvalue is given for a) the sub-matrix including the  first three sectors and b) the original matrix.}
  \label{tab:Vordering_number_init}
\end{table}

\section{Three-partition Projection}

As discussed in the introduction, the central goal of the projection
techniques is to describe the low-lying generic physics of the
original quantum many-body system.  We have seen, that NPT as opposed
to standard projection techniques resulting in effective models, even
yields quantitatively good agreement with the result of the the
original model. If still higher accuracy is the ultimate goal, then
one can do even better then what has been discussed so far.  It is
possible to include in each recursion step two sectors at a time with
little more CPU effort.  To this end we consider a matrix which has
$3\times 3$ tridiagonal block structure, that results, if we add in
each recursion (projection) step two additional sectors to the first
partition. The matrix has the structure
\begin{equation*}
  \begin{pmatrix}
    A & E & 0 \\
    E^\dag & B & F\\
    0 & F^\dag & C
  \end{pmatrix}\;.
\end{equation*}
\begin{figure}[h]
  \centering
  \includegraphics[width=\linewidth]{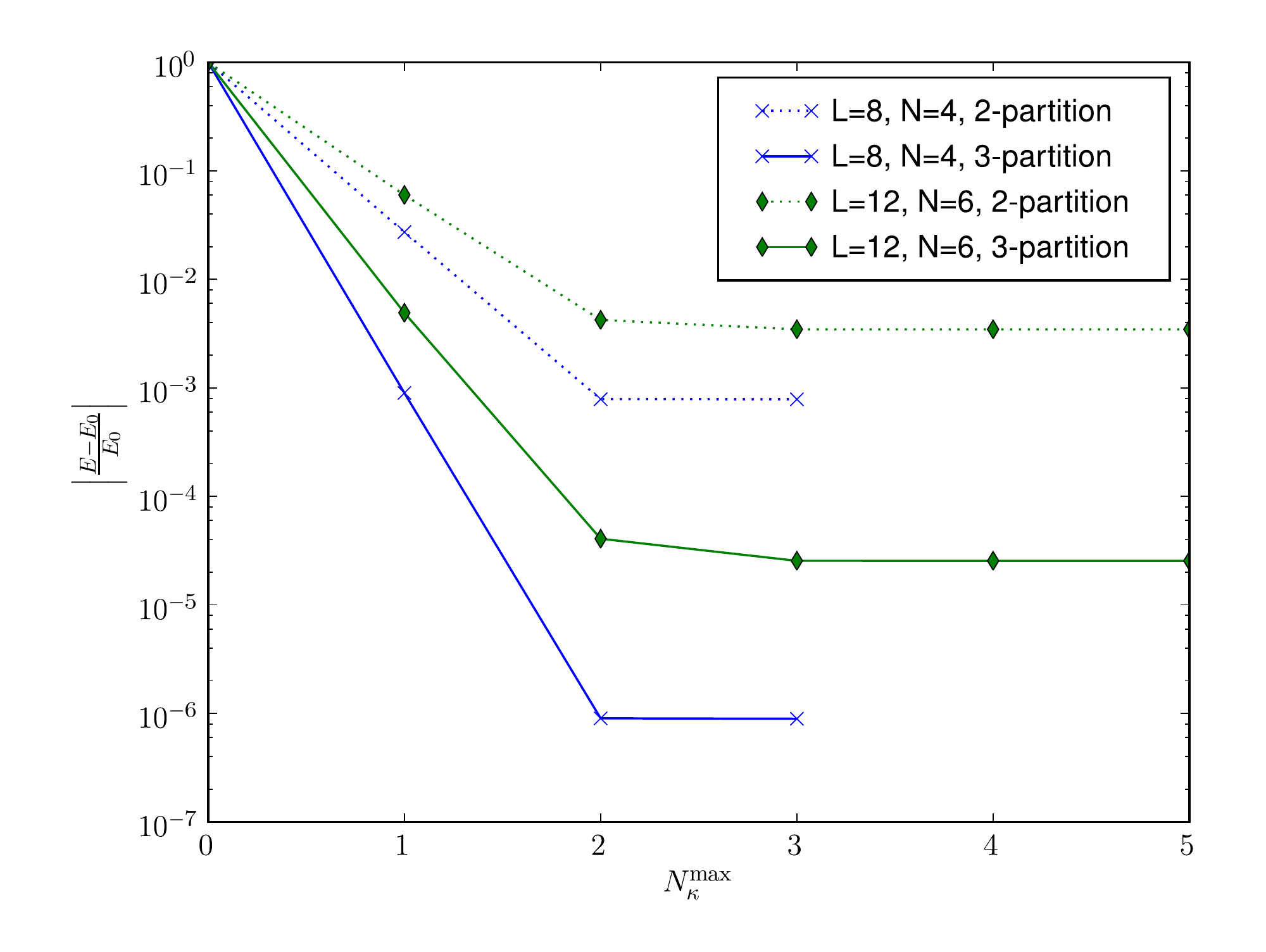}
  \caption{Comparison of the 2-partition and 3-partition
    approach for two different system sizes. The convergence of the relative uncertainty of the lowest
    eigenvalue is shown versus the maximum number of sectors included.}
  \label{fig:VORDERING_COMPARISON_2PART_3PART}
\end{figure}

Here, an equivalent relation to Eq. \ref{eq:Vordering_proj_x1} can be obtained:

\begin{equation*}
  \left(A - E\left(B-\lambda I-F(C-\lambda I)^{-1}F^\dag\right)^{-1}E^\dag\right)\vec x_1=\lambda x_1\;.
\end{equation*}

The contribution of the matrix $C$ shall be expanded up to the linear
inverse term:
\begin{align*}
  A-\lambda I - &E\left(B-\lambda I\right)^{-1}E^\dag \\
 - &E(B-\lambda I)^{-1} F(C-\lambda I)^{-1}F^\dag(B-\lambda I)^{-1}E^\dag
\end{align*}

Again, the expression is expanded in terms of the kinetic part:

\begin{multline}
  E(B-\lambda I)^{-1} F (C-\lambda I)^{-1} F^\dag (B-\lambda I)^{-1} E^\dag=\\
\sum_{\nu \mu \rho}\frac{1}{(B_V-\lambda)^{\nu+1}}\frac{1}{(B_V-\lambda)^{\mu+1}}\frac{1}{(C_V-\lambda)^{\rho+1}} \times\\
E(-\tilde B)^\nu F (-\tilde C)^\rho F^\dag (-\tilde B)^\mu E^\dag\;.
\end{multline}

Using this extension, the numerical result can be improved by orders
of magnitude as is shown in
Fig. \ref{fig:VORDERING_COMPARISON_2PART_3PART}. The results are for
the spinless fermion model at strong coupling $|V/t|=10$ and
half-filling. In both case the initial partition is given by the
lowest sector without further truncation, i.e. $p_1=\tilde p_1 = 2$.

\section{Results for the Hubbard model}

Next we present some results for the Hubbard model.
\begin{table}[h]
  \centering \footnotesize
 \begin{tabular}{cccc}
\hline
\hline
    system & model & $E_\text{NPT}$ & $\left|\frac{E-E_0}{E_0}\right|$\\
\hline
    \multirow{3}{0.4\linewidth}{$L=8$, $N_\spinup=N_\spindown=4$, $U=10$}& Hubbard & -2.1767\\
    &NPT ($\tilde p_1=20$)& -2.1526& 0.0110\\
    &Heisenberg & -2.2604& 0.0385\\
\hline
\multirow{3}{0.4\linewidth}{$L=10$, $N_\spinup=N_\spindown=5$, $U=10$}& Hubbard & -2.7037\\
&NPT ($\tilde p_1=26$)& -2.6572& 0.0172 \\
&Heisenberg&-2.8062& 0.0379\\
\hline
\multirow{3}{0.4\linewidth}{$L=10$, $N_\spinup=N_\spindown=4$, $U=10$}& Hubbard & -5.6698\\
&NPT ($\tilde p_1=80$) & -5.7303&0.0107\\
&tJ&-5.5282&0.0250\\
\hline
\multirow{3}{0.4\linewidth}{$L=10$, $N_\spinup=N_\spindown=5$, $U=5$}& Hubbard & -4.9334\\
&NPT ($\tilde p_1=26$)& -4.6032& 0.0669\\
&Heisenberg&-5.6123& 0.1376\\
\hline
\hline
  \end{tabular}
  \caption{Comparison of the lowest eigenvalue obtained by NPT with those of the Heisenberg or $tJ$-model and the exact result for the original Hubbard model. Periodic boundary conditions are assumed. Details of the NPT parameters are given in the text. One sectors were taken as initial partition ($N_\kappa^\text{ini}=1$).}
  \label{tab:Vordering_tJ_Heisenberg}
\end{table}
Standard projection technique leads in the case of half-filling to the
Heisenberg model and away from half-filling to the $tJ$-model
\cite{Fulde_ElectronCorrelations}. These models correspond in NPT to
the situation that the initial partition is the first sector with
$N_\kappa = 0$ (no double occupancies) and only the second sector is
mixed in the projection step. On top of that, the non-linear
eigenvalue problem is replaced by the linear one with $\lambda=0$ in
Eq. \ref{eq:Vordering_proj_x2}. In
Tab. \ref{tab:Vordering_tJ_Heisenberg} the lowest eigenvalues for
different system sizes and particle numbers are given. In the
half-filled case the NPT results are compared with the exact results
for the corresponding Heisenberg model and away from half-filling with
those of the $tJ$-model.  In NPT we use the first sector ($N_\kappa =
0$) as initial partition, restricted to $k=0$. For the truncation size
in the projection steps we used $\tilde p_1 =2$. The NPT recursion
steps are stopped at an absolute accuracy $\varepsilon_\lambda =
10^{-4}$.

We compare NPT results and those of the effective models with the
exact lowest eigenvalue of the Hubbard model. We see that even in the
strong coupling case ($|U/t|=10$), investigated here, NPT yields
quantitatively better results than the effective models.

\section{Conclusions}

The numerical projection technique (NPT), presented in this paper, is
a numerical generalization of standard projection technique, routinely
used to derive effective models for qualitative description of the
generic low-energy physics of strongly correlated systems. The
standard projection technique corresponds to an approximate evaluation
of second order perturbation theory in the inverse interaction
parameter.

NPT on the one hand a systematic application of the projection
techniques to any many-body Hamiltonian, irrespective of the
complexity of the original model, also in cases where the analytic
approach would lead to unmanageable effective models. On the other
hand it is also applicable for moderate coupling parameters, where
second order perturbation theory is not reliable enough.

We have demonstrated that NPT yields fairly accurate results with
little CPU-effort as compared to state-of-the-art exact
diagonalization. Furthermore, by solving the non-linear eigenvalue
problem and including higher order terms the approach yields
significantly better results than traditional projection technique.

\bibliographystyle{model1-num-names}
\bibliography{refs.bib}

\end{document}